# Rapport Technique

01 Septembre 2016

*Implémentation d'un Démon pour OpenBCI*

~

M. Chabance, G. Cattan, B. Maureille

Interactive human-machine technologies (IHMTEK)
Adresse : 30 Avenue du Général Leclerc, 38200, Vienne, France

**Résumé** – Nous décrivons ici une étude technique du casque d'électroencéphalographie (EEG) OpenBCI (New York, É.-U.). En comparaison au matériel de recherche, le casque d'OpenBCI est peu cher et donc susceptible de se populariser rapidement. Cette étude concerne la conception d'un *démon* (*i.e.,* un programme s'exécutant en arrière-plan et de façon continue), dont l'implémentation est facilement réutilisable, et ayant pour but d'acquérir et d'analyser les données EEG du casque. Cette étude a été réalisée par l'entreprise IHMTEK (Interaction Homme-Machine Technologie, Vienne, France) en 2016 dans le cadre d'une thèse CIFRE portant sur l'intégration, pour le grand public, d'une interface cerveau-machine basée sur l'EEG dans la réalité virtuelle.

**Abstract** – This document describes a technical study of the electroencephalographic (EEG) headset OpenBCI (New York, US). In comparison to research grade EEG, the OpenBCI headset is affordable thus suitable for the general public use. In this study we design a *daemon*, that is, a background and continuous task communicating with the headset, acquiring, filtering and analysing the EEG data. This study was promoted by the IHMTEK Company (Vienne, France) in 2016 within a thesis on the integration of EEG-based brain-computer interfaces in virtual reality for the general public.

**Introduction**

Les Interfaces Cerveaux-Machines (ICM) permettent à l'utilisateur d'interagir avec la machine sans passer par un intermédiaire mécanique. Elles ont un potentiel unique pour la Réalité Virtuelle (VR) en ce sens qu'elles diminuent la distance entre l'utilisateur et son avatar en remplaçant les moyens tangibles (souris, clavier) et audibles (assistants vocaux) d'interactions. Parmi ces ICM, celles reposant sur une stimulation visuelle occasionnelle ont l'avantage d'être robustes et peu fatigantes visuellement et mentalement en comparaison d'autres paradigmes (*e.g.*, (1)). Dans le cadre d'un sujet de thèse, la société IHMTEK étudie le couplage des ICM basées sur l'électroencéphalographie (EEG) avec la VR dans l'optique de rendre ces technologies accessibles auprès du grand public, pour un but de divertissement principalement. Il existe aujourd'hui un certain nombre de systèmes de VR à portée du grand public : SamsungGear (Samsung, Séoul, Corée du Sud), HTC Vive (HTC, Taoyuan, Taiwan et Valve, Bellevue, É.-U.), Oculus (Facebook, Menlo Park, É.-U.). Cependant, et jusqu'à récemment, le matériel EEG était à la fois cher et encombrant. Dans cette étude nous nous intéressons au casque OpenBCI (New York, É.-U.), un casque EEG possédant un prix abordable pour une petite structure voire un particulier - autrement dit, des qualités nécessaires pour populariser une technologie ICM+VR. Une étude précédente (2) a de plus montré que le matériel d'OpenBCI avait des performances proches, bien qu'inférieures, aux standards de recherche. Notons que la société Emotiv (San Francisco, É.-U.) propose également du matériel EEG accessible, et déjà étudié dans l'état de l'art (*e.g.*, (3,4)). Le casque d'OpenBCI s'accompagne également d'une plateforme d'analyse, mais dont les fonctionnalités nous apparaissent limitées pour les besoins d'une ICM. Par exemple, il n'est pas possible d'obtenir les informations affichées par la plateforme d'analyse depuis une application tierce. Également, il n'est pas clair comment étendre les fonctionnalités de cette plateforme pour donner accès à des informations haut-niveaux, tels que la présence du signal P300, un potentiel positif qui apparait dans le cerveau 240 à 600 ms après une stimulation visuelle, et dont la détection est difficile en raison de son faible rapport signal sur bruit. L'objectif de cette étude pilote est de développer un démon (*i.e.*, une tâche s'exécutant en arrière-plan et de façon continue) qui permettrait de commander le casque (démarrer, stopper ou mettre sur pause l'acquisition des données EEG ainsi que détecter des problèmes de connexion entre le casque et la plateforme d'analyse), exploiterait les données brutes et fournirait une interface pour d'autres applications dans le but de communiquer (*e.g.*, (5,6)) ou de se divertir (*e.g.*, (7,8)). Ces applications se connecteraient au démon qui permettrait aussi l'accès à des données haut-niveaux comme la détection du signal P300.

**Participants**

Cinq sujets de genre masculin entre 20 et 30 ans ont testé notre implémentation du driver, à différentes étapes de développement. Les sujets étaient tous volontaires et travaillaient au sein l'entreprise IHMTEK. Tous étaient en bonne santé. Ils s'agissaient à la fois de stagiaires artistes et développeurs, d'ingénieurs et de commerciaux. Les sujets étaient avertis des risques de l'expérience, principalement un risque d'allergie due à l'utilisation d'une pâte conductrice et du contact des électrodes avec la peau. Autrement, l'expérience ne comportait aucun risque ni pour le sujet ni pour l'expérimentateur. Nous avons obtenu le consentement de chaque sujet, et chaque sujet était libre de partir à n'importe quel moment de la procédure de test. Un critère de non-inclusion était la prise de psychotropes ou d'alcool, lesquels sont susceptible de modifier l'aspect des signaux mesurés. Toutefois, ce critère n'a jamais été vérifié.

**Matériel**

Nous avons utilisé la carte embarquée Cyton d'OpenBCI (New York, US), qui permet l'enregistrement simultané de huit électrodes 'Gold Cup' (OpenBCI, New York, US) - **Figure 1a**). Les électrodes 'Gold Cup' sont des électrodes humides. Bien que plus compliquées à nettoyer, les électrodes humides enregistrent un signal de meilleure qualité (9). Elles nécessitent l'utilisation d'une pâte conductrice qui assure la transmission du signal électrique entre la peau et les électrodes (**Figure 1b**). Comme pâte, nous avons utilisé un gel de la marque UNI'GEL (CERACARTA, Forly, Italie), qui est hypoallergénique et non-irritant. Il est également hydrosoluble, ce qui permet de nettoyer rapidement les électrodes avec de l'eau après utilisation. Le signal des électrodes était enregistré à une fréquence de 250 Hz et sans filtrage digital, puis transmis au PC grâce à un dongle Bluetooth (Bluetooth basse consommation v4.0). La carte embarquée était alimentée par des piles.

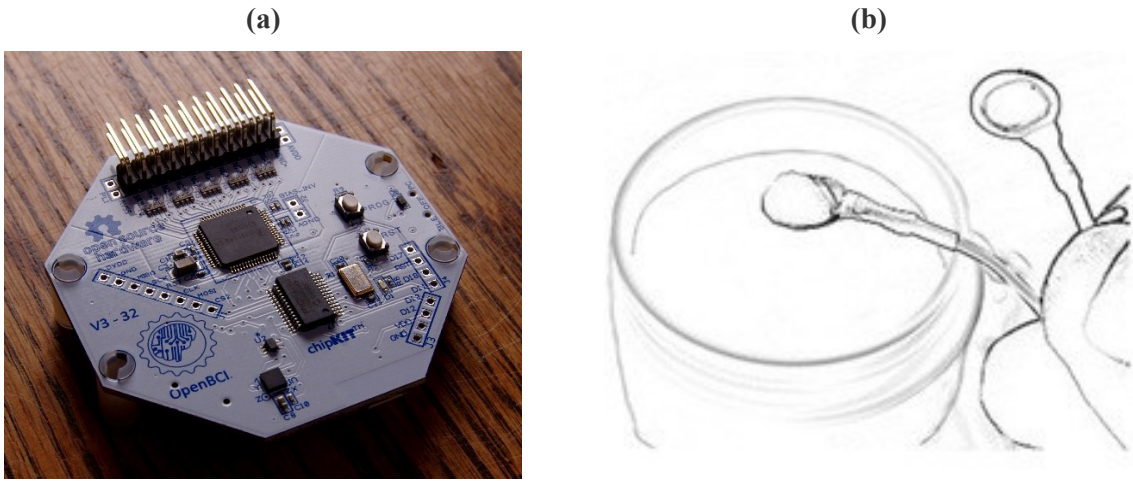

**Figure 1.** Images montrant : (a) la carte embarquée Cyton (OpenBCI, New York, US) et (b) les électrodes 'Gold Cup' qui se placent sur la peau ou le cuir chevelu grâce à une pâte conductrice. Source de l'image (a) : https://commons.wikimedia.org/.

**Implémentation**

Cette section décrit les composants logiciels et l'architecture matériels que nous avons mis en place pour implémenter notre démon. Le code est disponible en ligne sous l'adresse https://gitlab.com/ihmtek/openbci_daemon.git. Une autorisation peut être nécessaire pour accéder au répertoire.

*Composants logiciels*

Il existe des contraintes fortes quant aux composants utilisés pour implémenter notre démon. Dans l'optique d'une application grand public, le code développé doit être rapide et maintenable. Nous détaillons ci-dessous le choix des composants utilisés.

- *Langage de programmation*. IHMTEK possède une bonne connaissance du langage C++ et l'utilise dans d'autres projets. La force du C++ est sa performance, ce qui nous semblait nécessaire, car exploiter les données brutes du casque demande d'intenses calculs mathématiques (analyse du signal).

- *Interface graphique*. Qt (10) aurait été une très bonne candidate. Mais la bibliothèque est devenue payante pour les programmes non open source, et notre implémentation n'a pas forcément vocation à le devenir. *wxWidget* (11,12) reste le deuxième choix le plus populaire en C++, très complète et surtout portable.

- *Chaîne de compilation.* Pour garantir une compilation portable, nous utilisons l'outil *CMake* (13). Il permet de générer des makefile pour différentes plateformes. Ainsi nous compilons avec *gcc* sous Linux et le compilateur *msv* de Visual Studio sous Windows (Microsoft, Redmond, US). CMake possède une directive qui permet de trouver automatiquement les bibliothèques C++ les plus courantes (telles que wxWidget), ce qui simplifie le développement sous Windows.

- *Dialogue avec le port série.* Le casque OpenBCI est relié à l'ordinateur par l'intermédiaire d'un dongle Bluetooth branché par USB. Le pilote USB est en fait un convertisseur USB vers port série. L'envoi de commandes au casque se fait en écrivant des caractères ASCII dans ce port. Les données sont ensuite envoyées sous la forme d'un flux binaire dans ce même port. Le choix de la bibliothèque de lecture/écriture dans un port série se portait entre *Serial* (14), *LibSerial* (15) (synchrones et minimales) et *Boost.Asio* (16,17) (complète et asynchrone). Serial est plus simple et permet des lectures/écritures synchrones et avec des temps d'expiration. Boost.Asio force l'orientation asynchrone, ce qui complique l'architecture de l'application et entre en conflit avec le fonctionnement synchrone du dongle. Nous avons donc finalement choisi Serial. LibSerial semble offrir les mêmes fonctionnalités que Serial.

- *Interface pour les applications clientes.* Le choix s'est d'abord porté sur *websocket*++ (18). Mais lors de l'implémentation, cette bibliothèque semblait entrer en conflit avec Serial. Nous nous sommes rabattus sur Boost.Asio qui possède des sockets TCP. Cette fois l'approche asynchrone correspond bien au cas d'utilisation. Le principal inconvénient est la lourdeur de Boost, ainsi que ses obscurs messages d'erreurs dus à l'utilisation intensive des templates génériques en interne.

### *Architecture logicielle*

L'application est naturellement séparée en trois parties : l'interface graphique, la connexion avec le dongle, et la connexion avec les applications clientes (**Figure 2**). Ces trois parties possèdent leur propre thread. Les données brutes sont lues dans le thread côté dongle et y sont décodées en paquets. Quelques paquets disparaissent entre la carte et le programme, nous les approximons par interpolation linéaire. Ces paquets sont ensuite transmis au thread principal. Ce thread a pour tâche d'afficher les courbes dans l'interface graphique de l'outil et d'obtenir des informations haut-niveaux (ex. : détection du signal P300) utilisables par des applications ou des jeux. Ces informations haut-niveaux nécessitent des calculs mathématiques complexes (transformées de Fourier, réseau de neurones, etc.). Ces calculs n'ont pas encore été intégrés à l'application et sont testés sur Matlab (Mathworks, Natick, USA) ou OpenVibe (19,20). La bibliothèque Eigen (21) constitue également une alternative en C++. Lors de l'implémentation, il est préférable de déplacer les calculs gourmands dans un quatrième thread (les processeurs modernes ont pour la plupart quatre cœurs). Enfin, ces informations haut-niveaux sont mises à disposition des sockets qui l'enverront aux applications clientes. Ces sockets ont aussi le devoir de récupérer des informations nécessaires au traitement des données, comme le début des stimulations visuelles dans le cadre d'une application P300. La transmission de message d'un thread à un autre est principalement effectuée à l'aide des queues non bloquantes de Boost (*boost::lockfree::spsc_queue*), car un type d'information n'est fourni que par un thread et lu seulement par un autre (spsc signifie single-producer-single-consumer). Ces queues sont passées en argument au constructeur des objets.

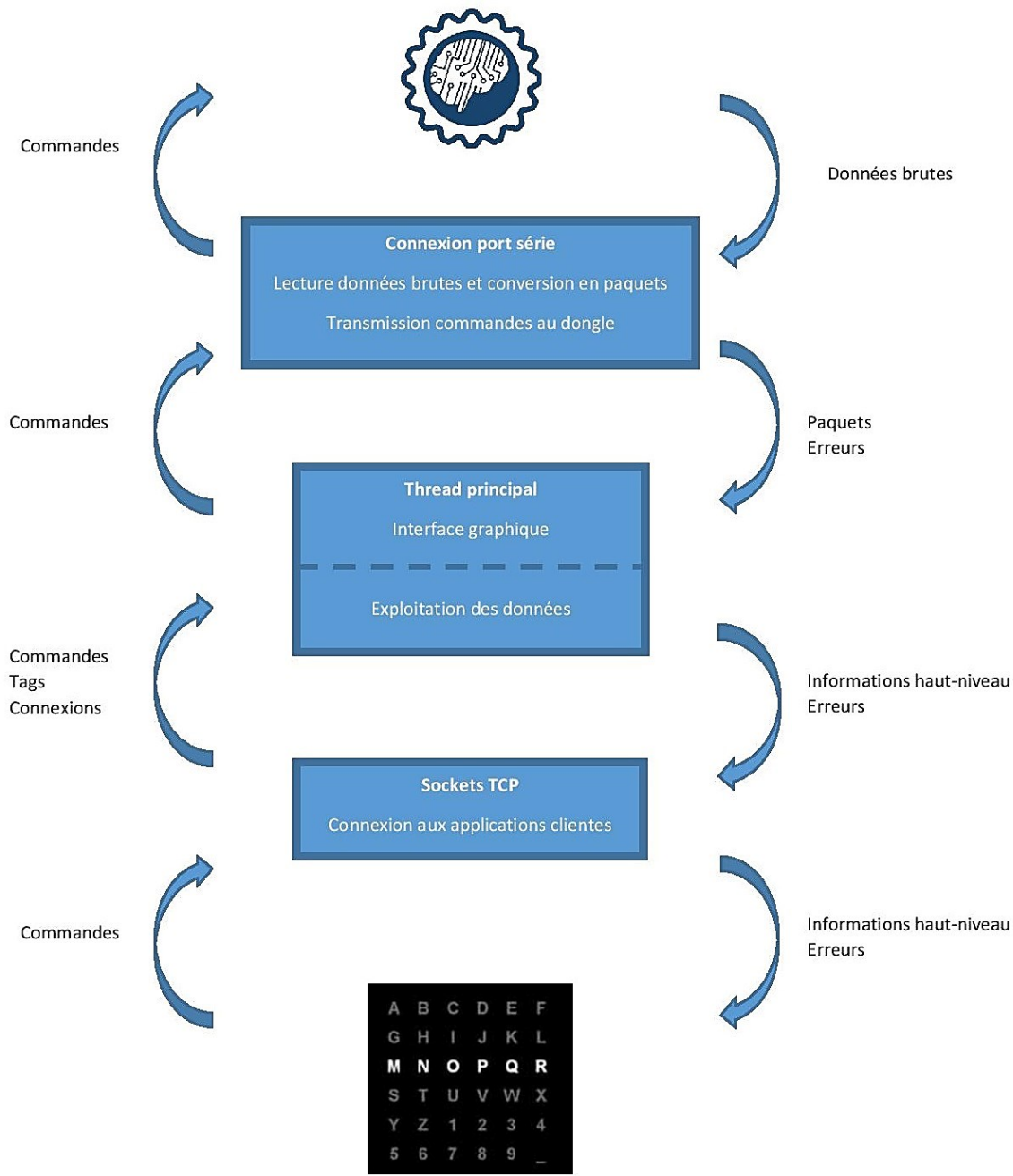

**Figure 2.** Architecture de l'application.

## Procédure

Le driver a été testé à différentes étapes de développement sur un à cinq sujets. Les tests ont eu lieu dans les locaux d'IHMTEK sur des personnes y travaillant. La pièce faisait environ 20m², et nous avons veillé à ce que les conditions d'aération et de luminosité soient suffisantes. Les locaux disposent également d'une salle d'eau, permettant de nettoyer le gel sur les participants et les électrodes entre chaque test. A cet effet, des produits nettoyants et des serviettes étaient laissés à disposition des sujets.

La procédure de test consistait à enregistrer et afficher simultanément le rythme cardiaque des sujets entre 10 secondes et une minute. Un opérateur mesurait manuellement le pouls du participant tout en regardant l'écran et s'assurait ainsi de la concordance entre le rythme cardiaque et le signal affiché. Le matériel d'OpenBCI s'accompagne également d'une plateforme d'analyse écrite en Processing (22) qui permet de visionner l'amplitude et la localisation du signal. Cette plateforme a permis également d'évaluer la qualité de notre driver en comparant le signal cardiaque enregistré par notre driver avec celui de la plateforme.

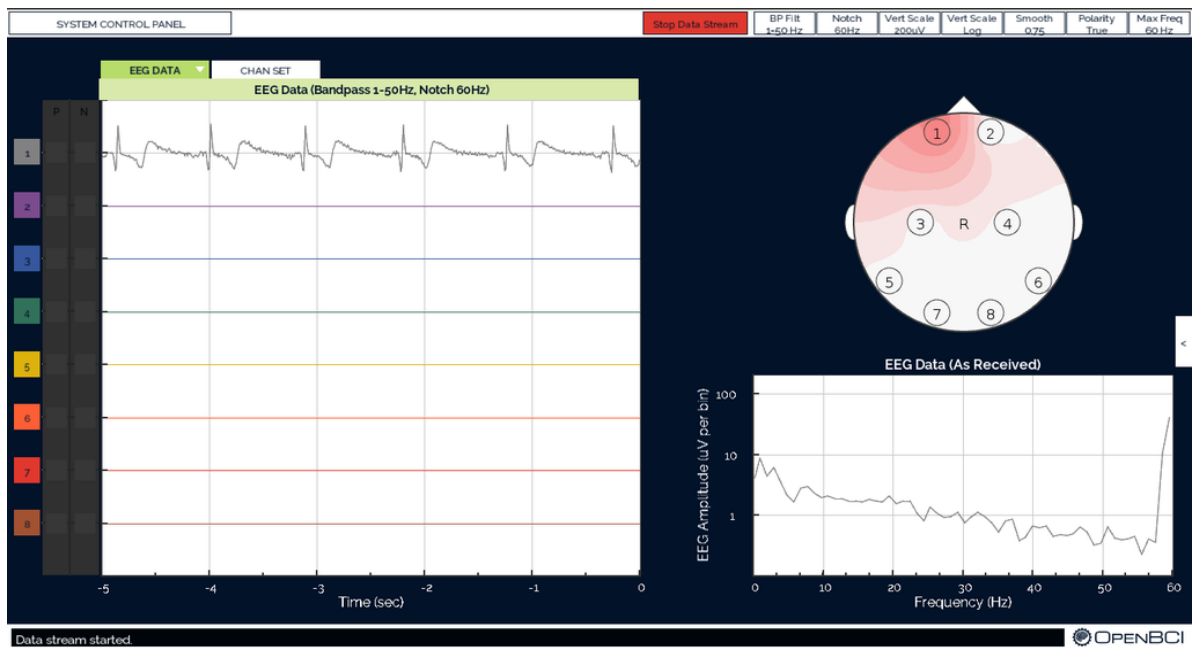

**Figure 3.** Capture d'écran du logiciel d'analyse d'OpenBCI. Le logiciel permet d'afficher le signal temporel (en bas à droite) d'une ou plusieurs électrodes et leurs localisations sur le scalpe (en haut à droite). Le graphique à gauche de l'écran affiche le signal filtré, ici avec un filtre passe-bande dans la région 1 à 50 Hz et un filtre Notch à 60Hz. Source : https://commons.wikimedia.org/.

Bien que le cadre de l'étude concerne les interfaces cerveau-ordinateur, l'implémentation que nous proposons a été testée sur un signal cardiaque, car celui-ci possède un haut rapport signal sur bruit (en comparaison au P300 par exemple) et peu facilement être mesuré par un opérateur humain. En outre, il suffit d'une électrode pour détecter ce signal, alors que l'enregistrement EEG nécessite de répartir plusieurs électrodes sur la surface du scalpe. Par exemple, le jeu *Brain Invaders* (23,24) utilise 16 électrodes pour détecter le signal P300.

**Discussion**

L'implémentation de notre driver donne des résultats en tout point conformes au logiciel d'OpenBCI et au rythme cardiaque mesurée par l'opérateur. Elle est également rapide et maintenable. L'approche par module que nous avons adoptée, et le soin porté dans le choix des bibliothèques le rendent modulable et réutilisable. Ce driver constitue donc une brique logicielle qu'IHMTEK ou d'autres développeurs peuvent s'approprier pour créer des produits personnalisés avec le matériel d'OpenBCI. Cependant, les tests avec les participants montrent que ce driver reste peu robuste et présente une instabilité résiduelle. Les principales limites semblent venir du matériel, et de la documentation qui n'est pas suffisamment détaillée :

- *Instabilité.* Le logiciel de présentation écrit en Processing (22) fourni par OpenBCI semble stable. Par contre, impossible d'obtenir le même résultat avec notre démon. Nous avons pourtant respecté les différentes initialisations ainsi que les délais entre chaque commande. A l'heure actuelle, allumer puis éteindre la lecture des données résulte une fois sur deux par la déconnexion du dongle. Nous ne savons pas si le problème déconnexion a lieu au niveau de la connexion USB ou entre le dongle et la carte d'acquisition.

- *Peu adapté au P300.* Dans un premier temps, nous comptions mettre en place la détection du P300 visuelle, un potentiel positif qui apparait sur l'EEG 240 à 600 ms après une stimulation visuelle. Malheureusement la détection de ce signal nécessite de connaître avec précision le début de chaque stimulation visuelle. D'autres systèmes (*e.g.,* (23)) utilisent un dispositif matériel pour cela, en *taguant* l'EEG au moyen d'un câble USB, au moment précis où une stimulation visuelle débute. Ces tags sont contenus dans une trigger channel qui se présente comme un canal d'écriture supplémentaire sur l'unité d'acquisition de l'EEG. Dans la mesure où le casque d'OpenBCI communique avec la plateforme d'analyse par Bluetooth, cette solution n'est pas envisageable dans notre cas. Une solution est de taguer logiciellement l'EEG. Les applications clientes enverraient au démon les tags et c'est lors de l'exploitation qu'elle les prendrait en compte. Dans l'étude (2), les auteurs mettent en place une application P300 fonctionnant avec OpenVibe et le matériel d'OpenBCI. Cependant, OpenVibe utilise un protocole de software tagging qui est peu robuste (24). La technologie LSL (25) est un protocole expérimental qui permet la gestion des aspects réseau et la synchronisation des flux des données. Elle pourrait permettre la synchronisation entre les tags et les données EEG. Les références (3,26–28) présentent des travaux qui montrent que la synchronisation logicielle entre les tags et l'EEG est possible. Par exemple, dans (3), les auteurs parviennent à détecter du signal P300 en utilisant un casque EEG sans fil et en extérieure. Elle est également utilisée dans (2) pour permettre la synchronisation entre OpenVibe et les stimulations envoyées depuis une application tierce en Python.

- *Une fréquence non normalisée.* La carte d'acquisition possède une fréquence de 250 Hz, ce qui n'est pas une puissance de deux, et ne constitue par une fréquence d'enregistrement standard. Nous devons modifier la fréquence d'échantillonnage pour 256Hz, ce qui résultera en une perte précision et en une augmentation des temps de calcul. L'*upsampling* est généralement obtenu par l'application d'un filtre dans l'espace fréquentiel.

- *Acquisition peu précise.* L'achat de la carte d'acquisition ne comprenait pas de structure pour placer les électrodes. Cette structure est vendue à part sous le nom de Ultracortex Marx IV (OpenBCI, New York, É.-U.) et est plutôt onéreuse (entre 300 et 400 euros). Nous avons essayé un bonnet artisanal composé de bandes élastiques. Le résultat n'est pas très concluant : les électrodes n'étaient certainement pas assez pressées contre le crâne. L'ajout de pâte conductrice n'a que très légèrement amélioré l'acquisition. La seule réussite a été la visualisation de fréquences alpha (ce sont des ondes à l'arrière du crâne qui sont facilement repérables et qui apparaissent notamment lorsqu'on ferme les yeux).

- *Peu de canaux.* La carte d'OpenBCI ne possède que 8 canaux. Ce qui est inférieur à la plupart des casques d'interface cerveau-machine. Il est possible d'y adjoindre une extension (le module Daisy, OpenBCI, New York, É.-U.) pour ajouter 8 canaux supplémentaires, mais au prix d'une division de la fréquence par deux. En effet le débit des données transférées de la carte à l'ordinateur ne change pas. A la place, un paquet sur deux de la carte principale est remplacé par un paquet du module Daisy. Ce problème lié à la fréquence d'échantillonnage du signal a aussi été souligné dans (2).

En résumé, nos efforts pour développer un driver personnalisé d'un casque OpenBCI se sont révélés infructueux. Pour des raisons logicielles, matérielles, mais aussi de documentations, il existe une instabilité et un manque de robustesse qui limite l'utilisation du casque à la commercialisation. Néanmoins, le matériel et le code développé restent appropriés dans un contexte de recherche ou pour des passionnés de l'informatique et de l'électronique. Nous soulignons également que le dispositif d'OpenBCI que nous avons testé est une première version d'un EEG à prix abordable pour des particuliers ou des petites entreprises, même si le cout de la structure supportant les électrodes reste globalement élevé. Nous fondons donc de grands espoirs quant à l'amélioration du produit dans les années à venir.

# Références

1. Sepulveda F. Brain-actuated Control of Robot Navigation. In: Advances in Robot Navigation [Internet]. Alejandra Barrera; 2011 [cited 2017 Aug 4]. Available from: http://www.intechopen.com/books/advances-in-robot-navigation/brain-actuated-control-of-robot-navigation

2. Frey J. Comparison of an open-hardware electroencephalography amplifier with medical grade device in brain-computer interface applications. ArXiv160602438 Cs [Internet]. 2016 Jun 8 [cited 2019 Mar 27]; Available from: http://arxiv.org/abs/1606.02438

3. Debener S, Minow F, Emkes R, Gandras K, de Vos M. How about taking a low-cost, small, and wireless EEG for a walk? Psychophysiology. 2012 Nov 1;49(11):1617–21.

4. Badcock NA, Mousikou P, Mahajan Y, de Lissa P, Thie J, McArthur G. Validation of the Emotiv EPOC® EEG gaming system for measuring research quality auditory ERPs. PeerJ. 2013 Feb 19;1:e38.

5. Käthner I, Kübler A, Halder S. Rapid P300 brain-computer interface communication with a head-mounted display. Front Neurosci. 2015;9:207.

6. Miralles F, Vargiu E, Dauwalder S, Solà M, Müller-Putz G, Wriessnegger SC, et al. Brain Computer Interface on Track to Home [Internet]. The Scientific World Journal. 2015 [cited 2018 Apr 29]. Available from: https://www.hindawi.com/journals/tswj/2015/623896/abs/

7. Marshall D, Coyle D, Wilson S, Callaghan M. Games, Gameplay, and BCI: The State of the Art. IEEE Trans Comput Intell AI Games. 2013 Jun;5(2):82–99.

8. Kaplan AY, Shishkin SL, Ganin IP, Basyul IA, Zhigalov AY. Adapting the P300-Based Brain #x2013;Computer Interface for Gaming: A Review. IEEE Trans Comput Intell AI Games. 2013 Jun;5(2):141–9.

9. Mayaud L, Cabanilles S, Langhenhove AV, Congedo M, Barachant A, Pouplin S, et al. Brain-computer interface for the communication of acute patients: a feasibility study and a randomized controlled trial comparing performance with healthy participants and a traditional assistive device. Brain-Comput Interfaces. 2016 Oct 1;3(4):197–215.

10. Rischpater R. Application Development with Qt Creator. Packt Publishing. 2014.

11. Smart J, Hock K, Csomor S. Cross-Platform GUI Programming with wxWidgets. Prentice Hall. 2005.

12. wxWidgets : A Cross-Platform GUI Library [Internet]. wxWidgets; 2019. Available from: https://www.wxwidgets.org/docs/

13. Martin K, Hoffman B. Mastering CMake [Internet]. Kitware, Inc. 2013. Available from: https://cmake.org/documentation/

14. Woodall W. Serial : A Cross-platform, Serial Port library written in C++ [Internet]. 2019. Available from: https://github.com/wjwwood/serial

15. CrayzeeWulf, Sauder M, Wedekind J. LibSerial : Serial Port Programming in C++ [Internet]. 2019. Available from: https://github.com/crayzeewulf/libserial

16. Schaeling B. The Boost C++ Libraries [Internet]. XML Press; 2014. Available from: https://theboostcpplibraries.com/

17. Torjo J. Boost.Asio C++ Network Programming. Packt Publishing; 2013.

18. Thorson P. C++ websocket client/server library. [Internet]. 2019. Available from: https://github.com/zaphoyd/websocketpp

19. Renard Y, Lotte F, Gibert G, Congedo M, Maby E, Delannoy V, et al. OpenViBE: An Open-Source Software Platform to Design, Test, and Use Brain–Computer Interfaces in Real and Virtual Environments. Presence Teleoperators Virtual Environ. 2010 Feb 1;19(1):35–53.

20. Arrouët C, Congedo M, Marvie J-E, Lamarche F, Lécuyer A, Arnaldi B. Open-ViBE: A Three Dimensional Platform for Real-Time Neuroscience. J Neurother. 2005 Jul 8;9(1):3–25.

21. Guennebaud G, Jacob B. Eigen [Internet]. 2010. Available from: http://eigen.tuxfamily.org

22. Reas C, Fry B. Make: Getting Started with Processing [Internet]. Maker Media. 2015. Available from: https://processing.org/reference/

23. Congedo M, Goyat M, Tarrin N, Ionescu G, Varnet L, Rivet B, et al. "Brain Invaders": a prototype of an open-source P300- based video game working with the OpenViBE platform. In: 5th International Brain-Computer Interface Conference 2011 (BCI 2011) [Internet]. 2011. p. 280–3. Available from: https://hal.archives-ouvertes.fr/hal-00641412/document

24. Andreev A, Barachant A, Lotte F, Congedo M. Recreational Applications of OpenViBE: Brain Invaders and Use-the-Force [Internet]. Vol. chap. 14. John Wiley ; Sons; 2016. Available from: https://hal.archives-ouvertes.fr/hal-01366873/document

25. Stenner T, Boulay C, Medine D. LabStreamingLayer [Internet]. Swartz Center for Computational Neuroscience; 2015. Available from: https://github.com/sccn/labstreaminglayer

26. Liao L-D, Chen C-Y, Wang I-J, Chen S-F, Li S-Y, Chen B-W, et al. Gaming control using a wearable and wireless EEG-based brain-computer interface device with novel dry foam-based sensors. J NeuroEngineering Rehabil. 2012 Jan 28;9:5.

27. Park J, Xu L, Sridhar V, Chi M, Cauwenberghs G. Wireless dry EEG for drowsiness detection. In: 2011 Annual International Conference of the IEEE Engineering in Medicine and Biology Society. 2011. p. 3298–301.

28. Sundararaman B, Buy U, Kshemkalyani AD. Clock synchronization for wireless sensor networks: a survey. Ad Hoc Netw. 2005 May 1;3(3):281–323.